# Prototyping a Serial Number based Authentication Model for a Computer in a Wireless Local Area Network

John C. Chebor, Simon M. Karume, Nelson B. Masese and Andrew Kipkebut

Kabarak University, School of Science, Engineering and Technology, Nakuru, Kenya

## Abstract

*With the increase of wireless LAN usage in homes and enterprises due to its numerous benefits, authenticating the ever increasing number of devices and their users has become a challenge to proprietors of such kind networks. A MAC address, a physical network address that is used as basis for this study, has a copy of its value in the system software that can be spoofed and altered rendering the address not unique, not secure and unreliable.  On the contrary, a computer's serial number is hard-coded in the system hardware only and therefore cannot be spoofed and altered making it unique, secure and reliable. The research, therefore, was aimed at designing a model that demonstrates how a computer's serial number can be used for authenticating a computer in a wireless local area network.  In order to achieve the research objective, the study examined the inbuilt access and use of a computer's serial number prototype model as an alternative method of authenticating devices in a network.  Design science research methodology that involved design and development, demonstration and model evaluation was employed.  A Serial Number Based Authentication prototype (SNAP) was therefore designed using state chart and flow chart diagrams based on dynamic programming, developed over evolutionary prototyping and test run on a static experimental design using Java Development Kit and MySQL platforms to demonstrate, as proof of concept, that a computer's serial number can be used to authenticate a computer in a wireless local area network. From the test runs whose outcomes were the binary values yes or no, it was found out that SNAP can actually allow or deny, enable or disable a computer in a network based on the computer's serial number.  The researcher therefore, recommends that the prototype be scaled up, then adopted as a network device authentication method*

## Keywords

*Computer's Serial Number, Authentication, Wireless LAN, Serial Number-Based Authentication*

## 1. Introduction

Wireless LANs (WLAN) also known as Wireless Fidelity (Wi-Fi) or 802.11 standards is a type of a local area network that allows users access network services using mobile devices (wireless stations) such as laptops, personal digital assistants, smart watches and even smartphones (Dordal, 2018).  The wireless stations use a base station usually an access point (AP) or a hotspot as an entry point to the network services (Romanov & Succi, 2018).  Unlike wired LANs that use cables or wires as transmission media, WLANs uses radio wave frequencies to transmit information over the local area network.

WLAN, therefore, comes with a myriad number of benefits as compared to wired LANs, notably, mobility, rapid deployment, reduction in infrastructure and operational costs, flexibility, and scalability (DHS, 2017 and Wallace, 2018).  Due to these benefits, hotspots are now virtually found everywhere; in enterprises, at homes, and in public places.  Wireless devices such as laptops, personal digital assistants and even smartphones come with Wi-Fi features integrated in them. Despite the numerous benefits that come with wireless LANs, network insecurity has become a thorn in the flesh for proprietors of such kind of networks. Singh & Sharma, (2017), points out that an attacker simply needs to be within the range of the WLAN access point to intrude into the network as opposed to wired LANs where an attacker requires physical access to the LAN or remotely use porous firewall systems to gain access to the LAN. WLANs are therefore easily targeted by attackers through spoofing, denial of service, eavesdropping, man-in-the-middle, masquerading, message modification, message replay and traffic analysis (Stallings, 2011; Poremba, 2017; DHS, 2017). Wallace, (2018), describes the reasons for the threats as default configurations, network architecture nature, encryption weaknesses, and physical security.

The numerous benefits that come with WLAN have made enterprises to adopt bring your own device (BYOD) concept to allow employees use their own devices (Poremba, 2017). Apart from the employees accessing and using enterprise networks, another group of stake holders such as visitors, vendors, and contractors at one point, if not all, can as well require network usage as they carry on their businesses with the enterprise. Rise in internet of things (IOT) devices such smart devices, smart watches and smart phones, as well, further complicates WLAN challenges equation (Pierce, 2021; Elkhodr & Mufti, 2019; Elkhodr et al. 2016). Allowing users to connect to the network with their own devices can pause as a security challenge as it becomes difficult for network administrators to control such kind of a network access and usage. It is therefore imperative that network administrators use network access control tools to control who should and who should not access the network. One of such kind of a control tool is the network access control (NAC) (Robb, 2022; Pierce, 2021; & Chiradeep, 2021). NAC constitutes identification, authentication, authorization, and accounting (IAAA) according to Lawson, (2017) as the essential functions in providing the required services in a network.

Major authentication methods or technologies began way back during the Second World War by the use of identification of a friend or a foe (IFF) (Lehtonen et al., 2008). From then on, advances in authentication techniques that include password, smart card, biometric, certificate MAC address, IP address and multi-factor (MFA) based authentication methods took effect (Shacklett, 2021; Johnson, 2021 and Fredriksson, 2017). The methods are categorized based on what one is known for or knowledge based (password, PIN), what one has or possesses (token, certificate), who one is or inheritance (biometrics), where one is or location based or address based (MAC address, IP address) and when one is authenticating or time factor as well (Shacklett, 2021). Out of all the mentioned authentication methods, MAC address, IP address and at times certificates are machine based authentication methods (Fredriksson, 2017). A MAC address, a physical network address that is used as basis for this study, has a copy of its value in the system software that can be spoofed and altered rendering the address not unique, not secure and unreliable, making it not suitable for authentication. A computer's serial number, in contrast, is only hard-coded on the hardware without a copy in the system software alone renders it hard to be spoofed (Derekyoung, 2017).

This research, therefore, was aimed at designing a model that demonstrates how a computer's serial number can be used for authenticating a computer in a wireless local area network through answers to the research questions: (i) How can an algorithm that can obtain and use a remote computer's serial number in a wireless LAN be developed? (ii) How can a model that uses the computer's serial number to authenticate the computer in a wireless LAN be designed? (ii) How can the model that uses the computer's serial number to authenticate the computer in a wireless LAN be demonstrated? (iv) How can the model that uses the computer's serial number to authenticate the computer in a wireless LAN be evaluated? The paper was then structured as follows: Section 1 provides background information and study motivation. Section 2 deals with related work. Section 3 presents the design, development, demonstration and evaluation of the prototype perceived to address the study questions listed. Section 4 presents the conclusion derived from the study

## 2. RELATED WORK

As a subset of location-based authentication, IP address authentication is a traditional method of authenticating computers that require network and resource access. Once a user logs onto a network, IP address authentication checks on their IP address and validates them against a list of allowed IPs or IP ranges. When a range of addresses is specified, the access point performs a logical and with the IP address entered in the IP address filter and the configured subnet. If an exact IP address is specified, the authentication method specifies a subnet mask so that only request from a client IP address is allowed or blocked, depending on what is configured in the filter (Servicenow, 2022). This eliminates or reduces the use of user IDs and passwords, initial configuration and maintenance is simple, works well with static IPs, however, requires a separate remote authentication tool, and slow for dynamic IPs, (EBSCO, 2022). But with dynamic nature of networks, that is, users and devices are mobile, plus users can use multiple devices from different locations, thus IPs might not correspond to their institutions. Although virtual private networks and proxy services have been fronted to remedy the drawback, they are complicated to manage have their own issues (Hoy, 2019). IP address range makes it easy to be spoofed, dynamic IP address allocation results to multiple users using the same IP, and the IP address keeps changing for the same device from location to location (Chad, 2017).

MAC address authentication, on the other hand, a port-based authentication method, allows or denies network access based on the MAC address credentials for machines such as IP phones, printers, and network attached storage devices. As a layer 2 OSI reference model issue, MAC address authentication solution uses RADIUS over IEEE802.1x framework rather than EAP (Cisco, 2018; Fredrisson, 2017). When a device connects to an access point (AP), the AP forwards the MAC address as the log in credential to the RADIUS server. With MAC-based authentication, the MAC address serves as both the username and the password. The RADIUS server consults the authentication server and sends back a RADIUS return attribute based on authentication results

MAC address filtering, also referred to as address control or address reservation, or still wireless MAC authentication allows or blocks traffic from a known machine or device depending on an organizational security policy to secure their networks (Andysah, 2017). However, due to a copy of a MAC address value in the system software (Cardenas, 2003), MAC addresses can be spoofed (Lee, 2010) and can be altered (Apple, 2011). Additionally, MAC addresses are not encrypted (Gill & Dahiya, 2017), it is not recommended for large wireless networks (Singh & Sharma, 2015; Watanabe et al, 2013) and furthermore, Kurose & Ross, (2013), raises the issue that a device can be attached to multiple networks each with a corresponding MAC address interface. A device, for instance can be attached to an Ethernet port, a Wi-Fi, or a Bluetooth port, all of which cannot uniquely identify the device. A MAC address in a nutshell, is not unique, is not secure and unreliable as required characteristics for identifiers (Developer, 2022)

## 3. SERIAL NUMBER BASED AUTHENTICATION PROTOTYPE

The serial number based authentication prototype (SNAP) that was perceived to address the challenges of using MAC address authentication, for authenticating computers in a network. The model was designed, developed, demonstrated and evaluated based on design science research (DSR) as follows

## 3.1. Design and Development

As earlier stated, the ultimate goal of the study was to design a model that demonstrates how a computer's serial number, as an identifier, can be used to authenticate a computer in wireless LAN. As such, authentication details that included the serial number, the IP address and the computer's name had to eventually be displayed on a display interface for further manipulation or be controlled if need arises. To achieve this ultimate goal, the SNAP model was designed and developed to first, registers computers so that it can later be used as a yard stick to either allow or deny computers access to the network. Then upon executing the application, it deletes existing authentication details in order to pave way for newly logged on computers' authentication details. This is followed by system collecting IP addresses, computer names, and computers' serial numbers for all computers that log on to the network. The authentication details are then collated for the purpose of validating them, as a unit, based on their registration status. Registered computers are allowed access to the network and unregistered computers are denied network access. Once they are allowed network access, valid computer authentication details are displayed on a Connected Devices display interface. Apart from containing computer's Serial Number, computer's Name and computer's Ip Address, the Connected Devices display interface contains an additional Status column to allow the validly logged on computers to be controlled if need arises. All the fundamental components, their details and relationships are illustrated in the figure 1 below

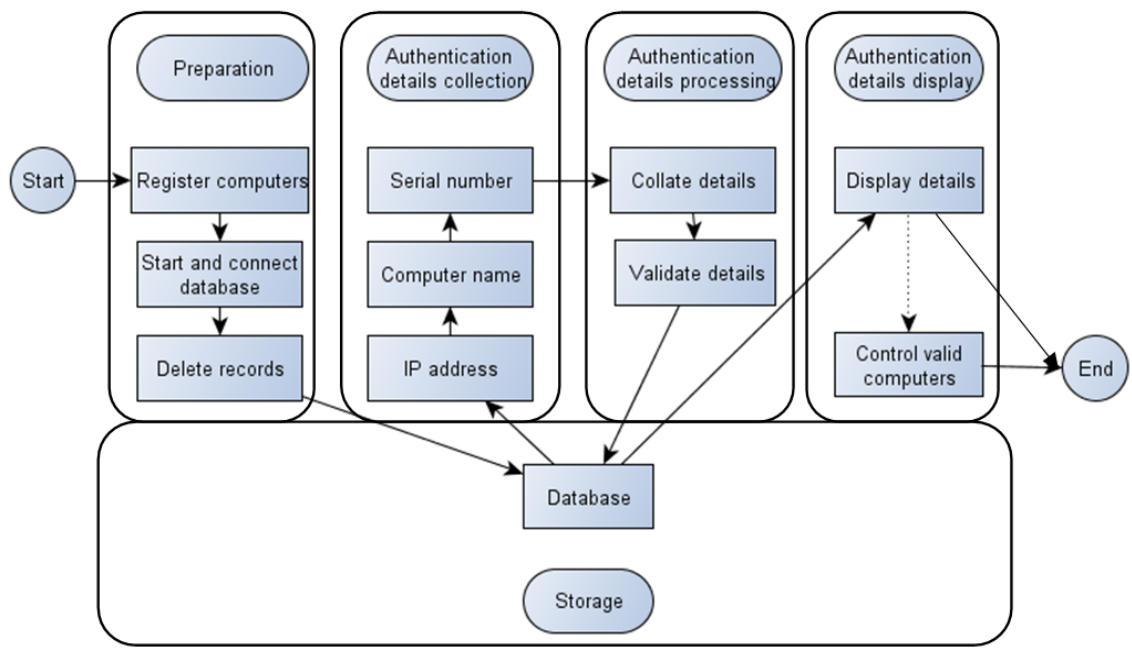

Figure 1. General model for a serial number based authentication system

A dynamic programming design was adopted during the algorithm design development phase of the study due to the fact that a dynamic programming algorithm optimizes solutions on a step by step basis recursively to the whole (SmartDraw, 2018; Visual-paradigm, 2018; Paramalways, 2009). In other words, the results of one step solve the problem of another consecutive step recursively. The algorithm with a corresponding flow chart (figure 2) that depicts the system description is as follows:

1. Start
2. Register computers

International Journal of Wireless & Mobile Networks (IJWMN), Vol.14, No.4, August 2022

3. Delete existing connected computers details
4. Get connected computer details
   4.1 Get the computer name
   4.2 Get computer raw IP address
   4.3 Get computer serial number
   4.4 Collate computer IP address, name and serial number
5. If computer is registered
   5.1 Allow computer connect to the network
   5.2 Post Connected Computer Details to the Database
   5.3 Retrieve and Display the Connected Computer Details from Database
   5.4 Control validly allowed computers
6. Else, deny computer network access
7. End

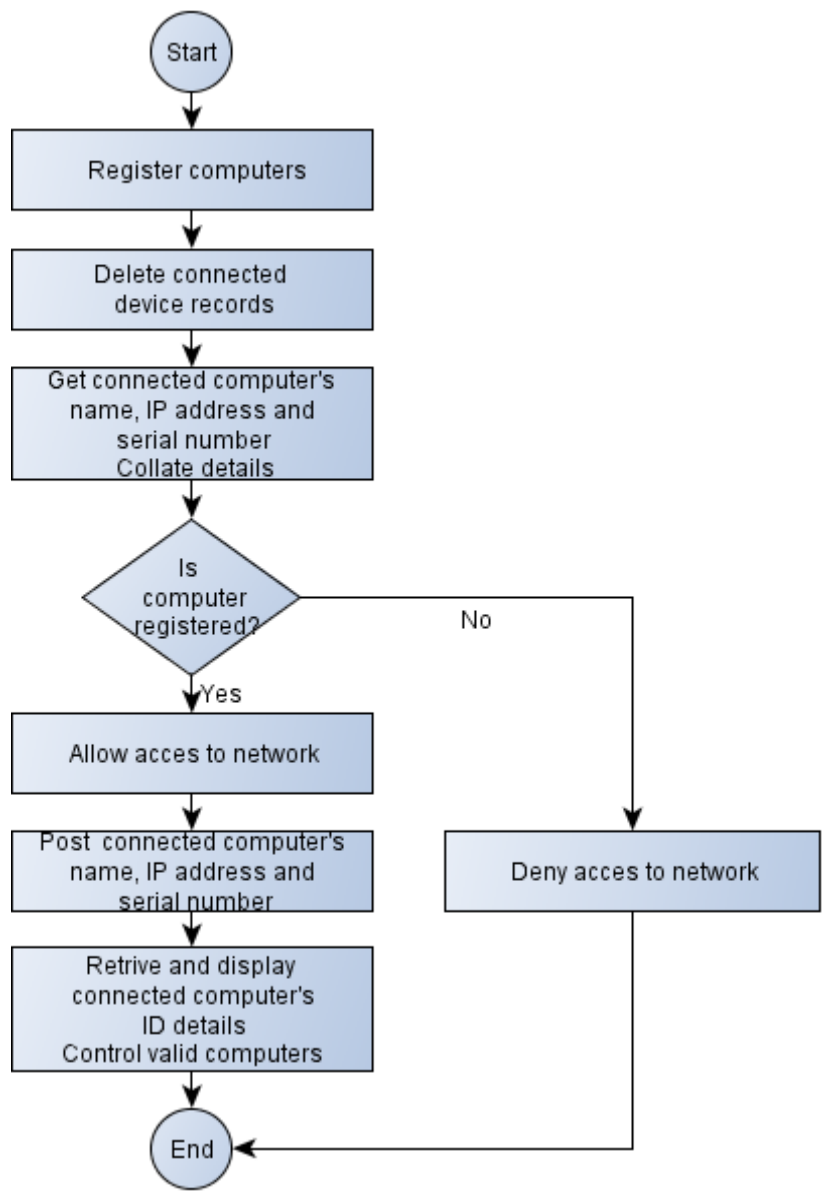

Figure 2. General flow chart for SNAM system

## 3.2. Demonstration

In order to demonstrate the proof of concept that a computer's serial number can be used to authenticate a computer in a wireless LAN, the SNAP model was implemented using MySQL database and Java's IDE tools over evolutionary prototyping using a static group comparison pre-experimental design set up. The set contained authentication server, authentication details database, an access point, two clients, the SNAP application and their connections as depicted in figure 3 below.

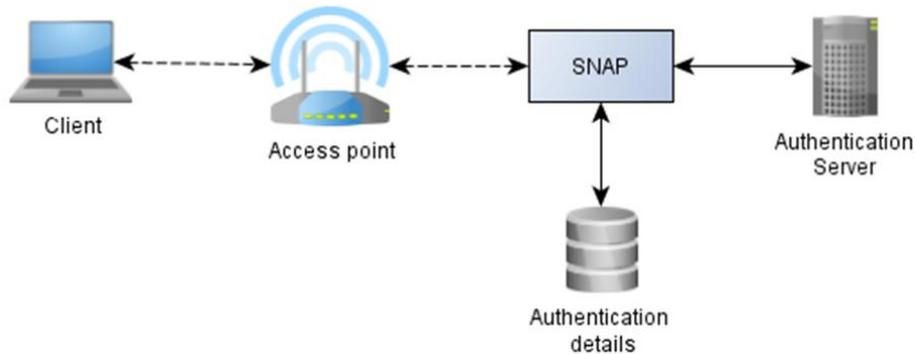

Figure 3: The SNAP system components

While the authentication server was installed with the SNAP application to perform authentication processing with whose details are stored and managed in the authentication database, the access point acted as a link between the clients and the server just as the case in a client-server architecture. Two clients were part of the set up so that one was configured for valid and the other for invalid expected outcomes. Apart from the need of using proof of concept (PoC) to prove the overall concept that a computer's serial number can be used to authenticate a computer in a wireless LAN, the set up as well was geared towards using PoC to prove that other modules of prototype that culminates to the overall concept, can as well be executed. PoC according to Leurs& Duggan, (2018) and MacPherson, (2018) is an exercise to test design or assumption ideas. The other concepts of the study, that correspond to the system modules, and required PoC were that;

1. A computer's authentication details (that is name, IP address and Serial number) can be collected
2. A registered computer can access a network
3. An unregistered computer cannot access a network
4. An already logged on or an allowed computer can be denied access to the network if a need arises

To begin with, four computers were registered as displayed in the `Registered Devices` interface shown the figure 4 below

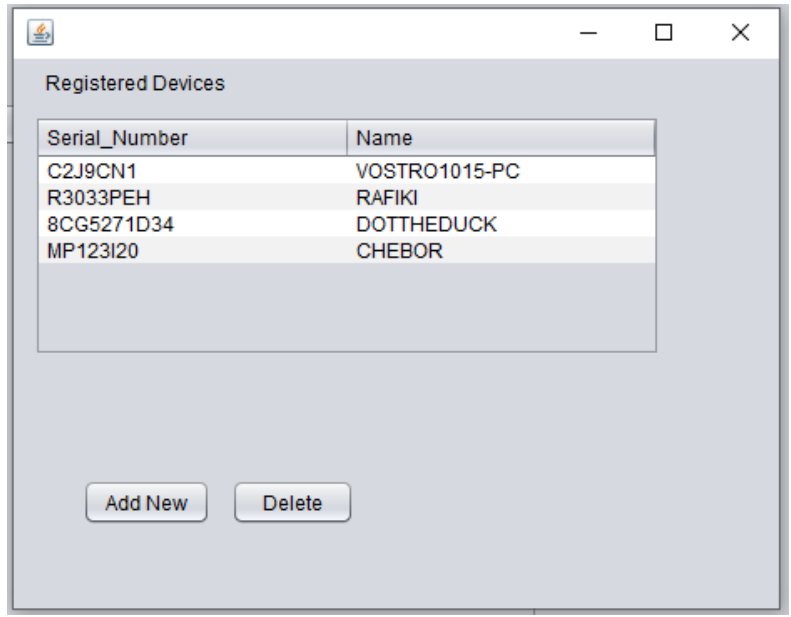

Figure 4. Computers that are already registered in the system

During the demonstration, three computers, `KOLSOLT` (unregistered client), `RAFIKI` (registered client) and `DOTTHEDUCK` (server) were connected. On test running the SNAP application, only `RAFIKI` and `DOTTHEDUCK` computers were allowed access while the unregistered `KOLSOLT` computer was denied network access as illustrated by the disabled state of the `Status` column on the `Connected Devices` display diagram in the figure 5 below

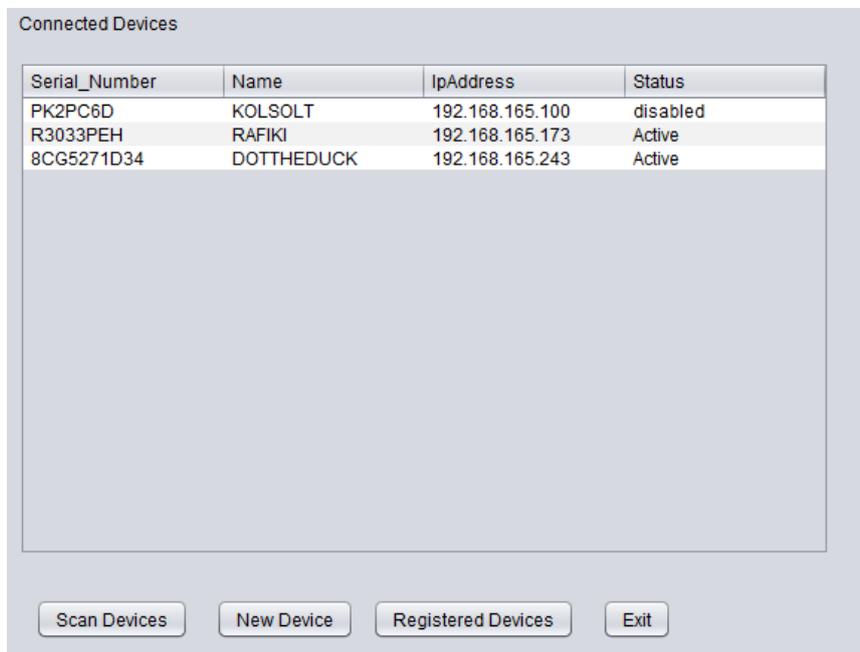

Figure 5. Blocked unregistered `KOLSOLT` computer

The Wi-Fi interface status for the `KOLSOLT` was equally disabled as indicated in the figure 6 below

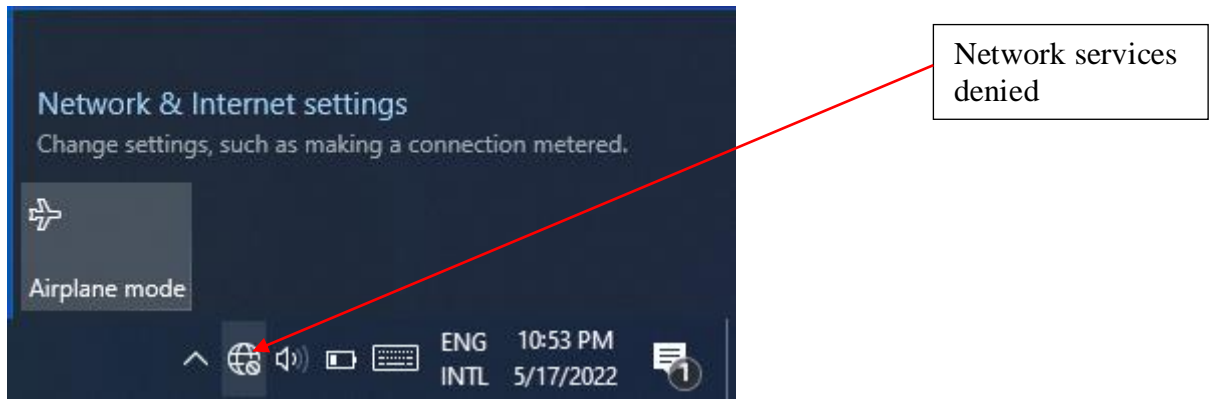

Figure 6. Blocked unregistered `KOLSOLT` computer Wi-Fi interface status

But once `KOLSOLT` was registered and the SNAP application executed once again, `KOLSOLT` was allowed network access together with other two registered computers as shown in the figure 7 below

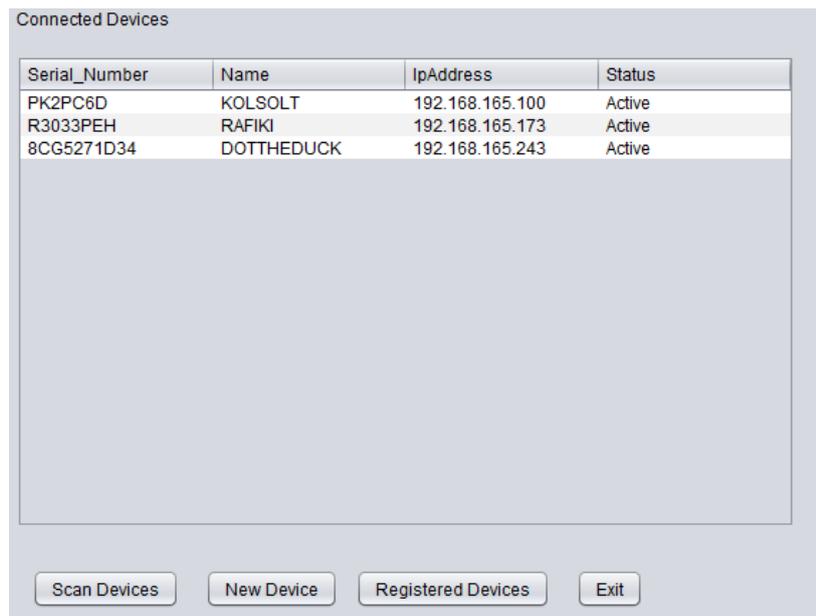

Figure 7. KOLSOLT allowed access after registration

The other aspect of the prototype was to develop and test run a section that controls validly allowed computers if need arises, while allowing other computers to continually use the network. This was achieved by first clicking on the desired client (e.g RAFIKI client computer in this case) on the Connected Devices interface on figure 7 above.

The system prompts on the surety to disable the client from accessing the network figure 8

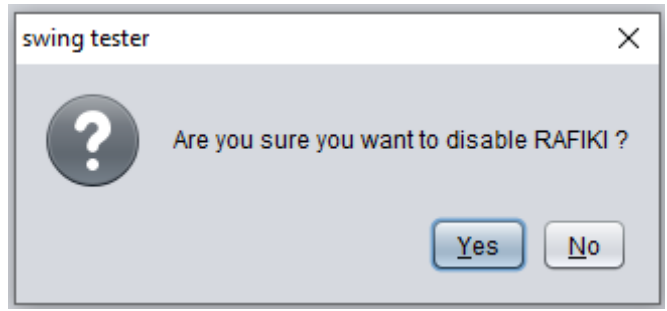

Figure 8. Denying RAFIKI computer network access prompt message

On clicking YES button, the RAFIKI client is successfully denied network access as illustrated in system prompt in figure 9 below.

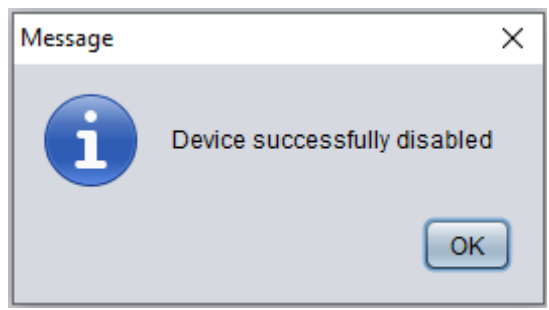

Figure 9. Denying `RAFIKI` computer network access confirmation message

The confirmation that `RAFIKI` client has been disabled, therefore, cannot access the network is illustrated in a screen shot in figure 10 below

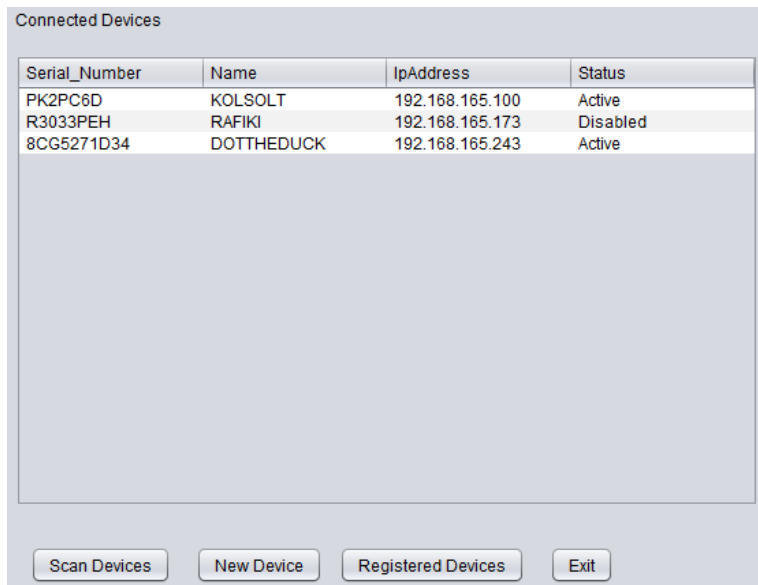

Figure 10. Denied `RAFIKI` computer network access

The client Wi-Fi interface status as well indicates that the client cannot access the network as shown in the figure 11 below

35

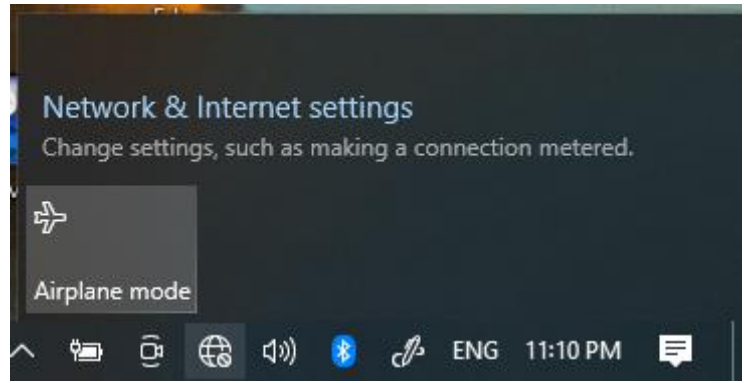

Figure 11. Blocked network denied `KOLSOLT` computer Wi-Fi interface status

A summary of the test runs on whether the prototype can or cannot register computers, collect computers authentication details, allow or deny a computers network access based on whether they are registered or not, and allow a valid computer to either continue using the network or be denied network resource usage due one or another reason were summarized as in table 1 below.

Table 1: System modules execution report

| Functions | Description | Execution Status (Yes/No) | Remarks |
|---|---|---|---|
| New Computer registration | Check if a new computer can be registered | Yes | Are confirmed in the Registered Devices interface |
| Collect Computer's Serial Number, IP address, Name | Check if authentication details can be collected from the system | Yes | Collected identification details are displayed on the Connected Devices interface |
| Collect Computer's authentication details with different access points (APs) | Check if authentication details can be collected from the system using different APs | Yes | Collected authentication details are displayed on the Connected Devices interface for three different APs |
| Allow a registered computer network access | Check if a registered computer is allowed network access | Yes | Registered computers are displayed in the Connected Devices interface |
| Deny an unregistered computer network access | Check if an unregistered computer is denied network access | Yes | Unregistered computers are denied network access as indicated in the affected computers network access status interface |
| Disable allowed computer | Check if an allowed computer can be disabled | Yes | This is confirmed by the disabled status of the computer in the Connected Devices interface and in the affected computers network access status interface |

### 3.3. Prototype Evaluation

For validation purposes, proof of concept method on the cumulative essential functionalities of the prototype was evaluated using test runs as proposed by Diceus, (2020) based on goal-based evaluation method (Cronholm & Goldkuhl, 2003). With the focus of being on intended services and outcomes of a program, goal-based evaluation measures the extent to which a program attains its intended goals. The prototype was fundamentally geared towards the ability to

1. Collect a computer's authentication details (that is name, IP address and Serial number)
2. Allow a registered computer access to a network
3. Deny an unregistered computer access to a network
4. Control an already logged on computer to a network if need arises

A set of three test runs were carried out using static group comparison pre-experimental design. Just as in the initial experimentation set up, each set had three computers and an access point. One of the computers, that was configured as server, was installed with the serial number authentication application. For comparisons purposes, the other two computers were configured as clients, one for valid and the other for invalid expected outcomes.

Each set was carried out independently by a wireless LAN stakeholder (that is, a network administrator, a Wi-Fi owner and an IT expert). Each participant was given instruction on how to run the system against a check list of the four test runs. They were expected to observe the behavior when the system is run and manipulated according to the fundamental questions that corresponds to the system functionalities and answer yes or no on the check list form given to them. The questions on the checklist that corresponded with the evaluation goals were as follows;

1. Can the system collect a logged on computer's authentication details (that is name, IP address and Serial number)?
2. Can the system allow a registered computer access to a network?
3. Can the system deny an unregistered computer access to a network?
4. Can the system deny an already logged on computer to a network if need arises?

The summary of the results from the three evaluators as indicated in appendix were summarized as in the table 2 below.

Table 2: Evaluators system modules execution report

| # | Evaluation | Evaluator 1 | Evaluator 2 | Evaluator 3 |
|---|---|---|---|---|
| | | Yes/No | | |
| 1. | Can the system collect a logged on computer's authentication details (that is name, IP address and Serial number)? | Yes | Yes | Yes |
| 2. | Can the system allow a registered computer access to a network? | Yes | Yes | Yes |
| 3. | Can the system deny an unregistered computer access to a network? | Yes | Yes | Yes |
| 4. | Can the system deny an already logged on computer to a network if need arises? | Yes | Yes | Yes |

## 4. CONCLUSION

From the discussions based on the design, development, demonstration and evaluation of a prototype that uses a computer's serial number to authenticate a computer in a wireless LAN, the following findings, that are actually geared towards answering the study questions, can be made thus (i) an algorithm that can obtain a remote computer's serial number in a wireless LAN can be developed using dynamic programming algorithm design, (ii) a model that uses the computer's serial number to authenticate the computer in a wireless LAN can be designed using a state chart and flowchart diagrams, (iii) a model that uses the computer's serial number to authenticate a computer in a wireless LAN can be demonstrated using a prototype on MySQL database and Java's NetBeans IDE tools and (iv) that the model that uses the computer's serial number to authenticate the computer in a wireless LAN can be evaluated using a goal-based evaluation method by independent evaluators.  Furthermore, the test runs from the prototype set up indicated that the SNAM prototype could collect a logged on computer's authentication details (that is name, IP address and Serial number), can allow a registered computer access to a network, can deny an unregistered computer access to a network and that it can deny an already logged on computer to a network if need arises.  In turn, a serial number can therefore, be used to authenticate a computer in a wireless LAN.  It is then recommended that the prototype be scaled up so that it can adopted an authentication method.